# The 'coin-through-the-rubber' trick: an elastically stabilized invagination

Fanlong Meng

*Institute of Theoretical Physics, Chinese Academy of Sciences*
Phone: 86-15001345869
flm@itp.ac.cn

Masao Doi

*Center for Soft Matter Physics and its Applications, Beihang University*
Phone: 86-1082321461
masao.doi @buaa.edu.cn

Zhongcan Ouyang

*Institute of Theoretical Physics, Chinese Academy of Sciences*
Phone: 86-62554467
oy@itp.ac.cn

Xiaoyu Zheng

*Department. of Mathematical Sciences, Kent State University*
Phone: 1-3306729089
zheng@math.kent.edu

Peter Palffy-Muhoray

*Liquid Crystal Institute, Kent State University*
Phone:1-3306722604
mpalffy@kent.edu

Abstract: A spectacular trick of close-up magicians involves the apparent passing of a coin through a rubber sheet. The magic is based on the unusual elastic response of a thin rubber sheet: the formation of an invagination, stabilized by friction and elasticity, which holds the coin. By pressing on the coin, the invagination becomes unstable, and the coin is released. We describe the deformation analytically using a simple Hookean description, and examine the stability of the invagination. We finally compare the prediction of the Hookean analysis with numerical solutions of the neo-Hookean model, and provide a brief commentary on the origins of the trick.

# 1. Introduction

One of the most impressive magic tricks which can be performed without any special skills on the part of the magician is the 'coin-through-the-rubber' trick (CTRT), where an ordinary coin appears to pass through a sheet of ordinary rubber. In its simplest form, the trick proceeds as follows. A thin rubber sheet, with a coin on top, is placed so as to cover the opening of an empty cup. The edges of the sheet are held against the sides of the cup either by the magician, or a rubber band. A member of the audience is then invited to push down on the coin with his/her finger. As he/she does so, the rubber stretches a little, and then, suddenly, the coin appears to go through the rubber, and falls to the bottom of the cup. This process is illustrated in Figs. 1 and 2.

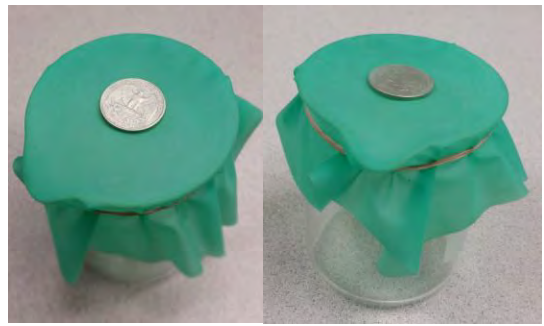

Fig. 1. Two perspectives of a latex rubber dental dam with a US quarter on top of a glass beaker.

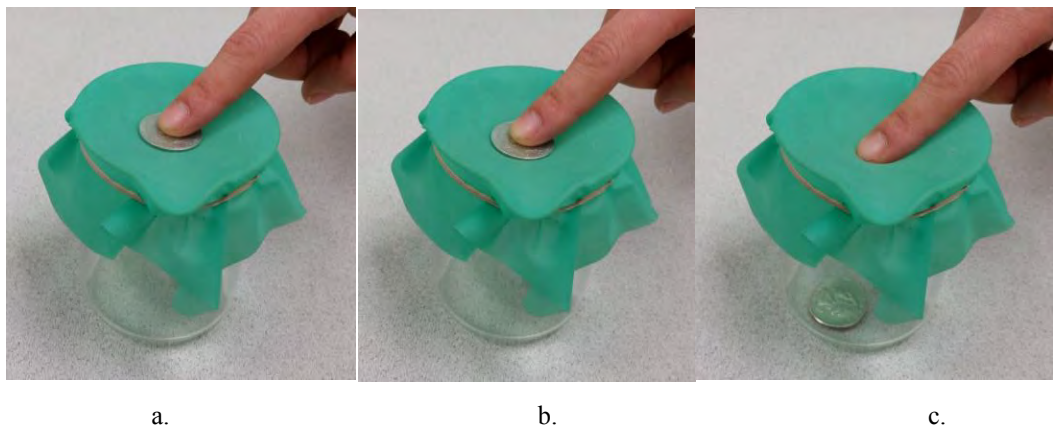

a.              b.              c.

Fig. 2. When pushed down, the coin 'magically' goes through the rubber, and falls to the bottom of the beaker.

The trick is remarkably effective, the phenomenon appears to be real magic. A brief history is given in the Appendix. Today, the CTRT is demonstrated on youtube [1]; a version is used to make smart phone covers using balloons [2]. We have also been able to perform the CTRT with a spherical glass marble instead of a coin, as well as with a variety of cylindrical objects.

The explanation of the magic is simple. The rubber is placed on top of the coin, and is stretched tightly; so tightly that the rubber becomes nearly transparent. This allows the illusion that the coin is on top of, rather than underneath, the rubber. Then, the stretched rubber is allowed to relax under the coin, contracting and forming an invagination which holds the coin in place below the rubber on top. This is illustrated in Fig. 3, showing the coin partially enseathed by the rubber from above, from the bottom and from the side.

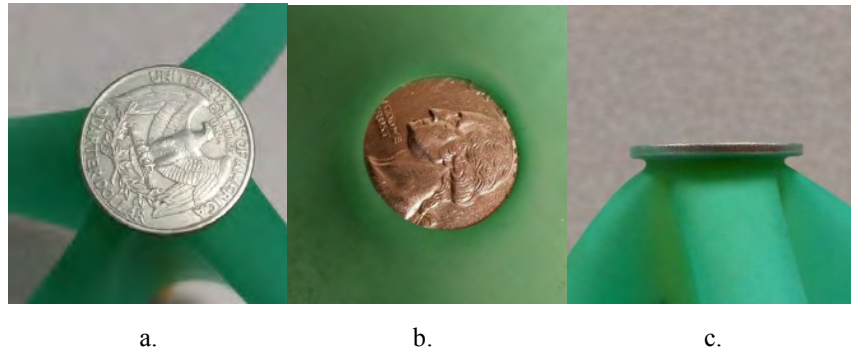

a.                 b.                 c.

Fig. 3 Images of the invaginated coin from above, from below and from the side.

The real magic is in the formation of the 'pocket'- of the invagination that holds the coin in place. The CTRT can be performed with latex sheets, such as dental dams, or condoms or balloons.

Invaginations in membranes are common in biology. They can be roughly divided into two classes. The first is where cell membranes partially envelop and entrap a variety of objects, ranging from prokaryotic organelles [3] to domain proteins [4]. In these instances, the invagination is driven by attractive surface forces [5], or curvature of the membrane [6-8]. Nonetheless, elasticity makes important contributions when certain proteins, such as clathrin, are involved in endocytosis [9]. Failed clathrin mediated endocytosis when, prior to scission, the contents of the vesicle – say a virus – is expelled, resembles somewhat the expulsion of the coin from the invagination in the rubber sheet. The second is where invagination occurs in groups of cells forming epithelial layers. Here elasticity and contractile stresses are thought to play important roles. Mechanical models have a long history [10]; a recent review [11] highlights current physical models of mesoderm invagination. Although we believe that the physics underlying the CTRT shares common features with invaginations in biology, establishing a direct connection is beyond the scope of this paper.

The phenomenon of the rubber partially enveloping the coin in CTRT is the first instance of an invagination stabilized solely by rubber elasticity and friction that we are aware of. Although there are some connections with cavitation[12], the phenomenon in CTRT is fundamentally different. The aim of this paper is to describe, in simple terms, the physics underlying elasticity stabilized invagination in elastic membranes.

## 2. Modeling the invagination

A wide variety of continuum models exist to describe rubber elasticity, where the strain energy density depends on invariants of the left Cauchy-Green stretch tensor [13]. One example is the general hyperelastic model proposed by Rivlin, where, for incompressible materials, the energy density has the form [14]

$$W_R = \sum_{p,q=0}^{\infty} C_{pq}(\lambda_1^2 + \lambda_2^2 + \lambda_3^2 - 3)^p (\lambda_1^2\lambda_2^2 + \lambda_1^2\lambda_3^2 + \lambda_2^2\lambda_3^2 - 3)^q, \tag{1}$$

where the $\lambda_i$'s are principal stretches, and, for incompressible materials, $\lambda_1^2\lambda_2^2\lambda_3^2 = 1$. The principal stretches are obtained from the eigenvalues of the stretch tensor [13]. The deformation can be described by $\mathbf{R}(\mathbf{r})$, where $\mathbf{r}$ is the position of a point in the undeformed rubber and $\mathbf{R}(\mathbf{r})$ is its position after the deformation. In our problem, due to symmetry and since the sheet is thin compared to its lateral dimensions, we assume that $\mathbf{R} = (R_r(r), 0, R_z(r,z))$, where $r$, $\phi$ and $z$ are the usual cylindrical coordinates.

If we assume further that $\partial R_z / \partial r$ is small, then the principal stretches are

$$\lambda_1 = \frac{\partial R_r}{\partial r}, \tag{2}$$

$$\lambda_2 = \frac{R_r}{r}, \tag{3}$$

and

$$\lambda_3 = \frac{\partial R_z}{\partial z}. \tag{4}$$

Rather than starting with an accurate non-linear model to understand how the invagination is stabilized, that is, to understand how the rubber grips the coin, we consider a simple Hookean model of the sheet. The advantage of this simple model is that it allows an analytic description of the deformation, and so gives ready insights into the underlying physics. In spite of its simplicity, the simple Hookean model gives the qualitatively correct behavior, as can be seen from comparisons with numerical solutions of the more realistic neo-Hookean model in Sect. 2.2.

### 2.1 The Hookean model

We begin our analysis by assuming small deformations, where $\lambda_i \simeq 1$. Writing $\lambda_3$ in terms of $\lambda_1$ and $\lambda_2$ and expanding the energy density of Eq. (1) in the vicinity of $\lambda_1 = 1$ and $\lambda_2 = 1$, we obtain, to lowest order,

$$W_R \simeq 4(C_{10} + C_{01})[(\lambda_1 - 1)^2 + (\lambda_2 - 1)^2 + (\lambda_1 - 1)(\lambda_2 - 1)], \tag{5}$$

where we take the $\lambda_i$'s to be positive [15]. Since the energy is now quadratic in the strains, this is the Hookean model; we believe all hyperelastic models reduce to this form in the limit of small strains. Because of its tractability, we examine the predictions of this Hookean model, even though typical strains in the CTRT are not small. In Sect. 2.2, we compare predictions of the Hookean model with those of the more realistic but less tractable neo-Hookean model.

We assume that our undeformed rubber sheet is a cylindrical disk of radius $r_m$ and thickness $h_0$. The total elastic energy, from Eq. (5), is

$$F = 4\pi G h_0 \int_0^{r_m} ((sR'-1)^2 + (\frac{R}{r}-1)^2 + (sR'-1)(\frac{R}{r}-1))rdr, \qquad (6)$$

where $G = 2(C_{10} + C_{01})$ [13] is the shear modulus. We have dropped the subscript $r$ on $R_r$, $R' = \partial R/\partial r$, and $s = sgn(R')$ needed to maintain the positivity of $\lambda_r$. The deformations of our rubber sheet are fully described by $R(r)$; the allowed deformations are those which minimize the energy $F$. In this description, we assume that the bending energy is negligible, and the sheet may be folded without any energy cost.

The Euler-Lagrange (E-L) equation is

$$rR'' + R' - \frac{R}{r} + \frac{3}{2}(1-s) = 0, \qquad (7)$$

which is linear, hence solutions which satisfy the boundary conditions are unique. The solution of the E-L equation is

$$R = Ar + \frac{B}{r} - \frac{3}{4}(1-s)r\ln r, \qquad (8)$$

and the constants of integration $A$ and $B$ are to be determined from boundary conditions.

It is useful to consider the tensile forces per length in the rubber sheet; these are

$$T_r = \sigma_{rr} h = T_0 s \frac{1}{2}\frac{r}{R}(2sR' + \frac{R}{r} - 3) \qquad (9)$$

in the radial, and

$$T_\phi = \sigma_{\phi\phi} h = T_0 \frac{1}{2}\frac{1}{sR'}(2\frac{R}{r} + sR' - 3) \qquad (10)$$

in the tangential directions, $\sigma_{rr}$ and $\sigma_{\phi\phi}$ are the radial and tangential stresses, $T_0 = 4Gh_0$, and

$$h = h_0 \lambda_z = \frac{r}{sR'R}h_0 \qquad (11)$$

is the thickness of the deformed rubber sheet. In equilibrium, the net force on any part of the rubber sheet must be zero. If the radial stress varies in the radial direction, the change must be balanced by the tangential stress, called, in this case, the hoop stress. On an element of the sheet shown in Fig. 4, the net radial force is $d(T_r R)d\phi$, while the net tangential force acting on the element in the radial direction is $T_\phi dR d\phi$.

Fig. 4. Schematic showing force balance via hoop stress in an annular region of the stretched rubber

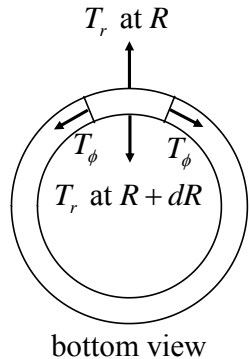

bottom view

Equating the net radial and net tangential forces in the radial direction gives

$$\frac{d(T_r R)}{dR} = T_\phi, \qquad (12)$$

which is just the condition that the divergence of the stress in the bulk vanishes. Substitution for the tensile forces per length gives Eq. (7); the E-L equation is the condition for force balance.

### 2.1.1 Solutions Describing the Deformation

It is convenient to work with dimensionless quantities, we therefore express all lengths in units of the coin radius $R_c$, and stresses in units of the modulus $G$. If the rubber is stretched so that points on a circle of radius $r_c$ in the unstretched sheet end up on the circumference of the coin, then $R(r_c) = 1$. The quantity $r_c$, whose reciprocal gives the stretch ratio of the rubber on top of the coin, is a key parameter of the problem.

The deformation is completely described by $R(r)$ of Eq. (8) where the constants of integration have been chosen to satisfy the boundary conditions. It is convenient to divide the rubber into three regions, as shown in Fig 5. Region 1 is on the top of the coin, as shown in Fig. 3a., where $0 \leq r \leq r_c$; here the rubber has been stretched, $R'$ is positive, and $s = 1$. Region 2 is underneath the coin, where the rubber is folded back, and $r_c \leq r \leq r_h$. We have denoted the edge of the hole, seen in Fig. 3b., in the unstretched rubber by $r_h$; here the rubber is stretched, but $R'$ is negative, and $s = -1$. Region 3 is the region $r_h \leq r \leq r_m$ extending from the hole to the edge of the rubber sheet; here the rubber may be unstretched, as shown in Fig. 3c., or stretched, as shown in Fig.2b.; in both cases, $s = 1$.

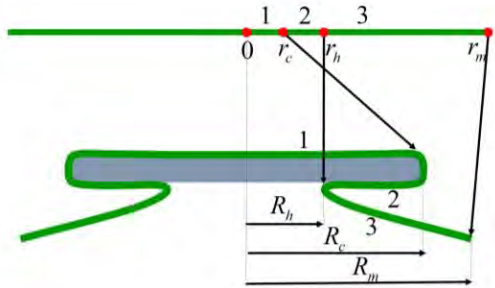

Fig. 5. Schematic of coin and rubber showing the three different regions.

Table 1 below summarizes the boundary conditions and the corresponding solutions.

| Region | s | BCs | Deformation | Comment |
|---|---|---|---|---|
| 1 | +1 | $R_1(0) = 0$<br>$R_1(r_c) = 1$ | $R_1 = \dfrac{r}{r_c}$ | $A_1 = \dfrac{1}{r_c}$, $B_1 = 0$ |
| 2 | −1 | $R_2(r_c) = 1$<br>$T_{r2}(r_h) = 0$ | $R_2 = A_2 r + \dfrac{B_2}{r} - \dfrac{3}{2} r \ln r$ | $r_h$ determined from $R_2(r_h) = R_3(r_h)$ |
| 3 | +1 | $T_{r3}(r_h) = 0$<br>$T_{r3}(r_m) = T_{ext}$ | $R_3 = A_3 r + \dfrac{B_3}{r}$ | $r_h$ determined from $R_2(r_h) = R_3(r_h)$ |

In each region, the constants of integration in the solutions describing the deformation can be determined from boundary conditions listed in Table 1. In regions 2 and 3, the hole radius $r_h$ is not known *a priori*; it is to be determined from the continuity condition $R_2(r_h) = R_3(r_h)$. Pushing down on the coin in the CTRT is equivalent to applying a tension per length $T_{ext}$ at the edge $r = r_m$ of the sheet. So if $T_{ext}$ is given, $r_h$ can be determined; conversely, if $r_h$ is given, $T_{ext}$ can be determined.

If there is no external tension applied at the edge, then the rubber in region 3 is undeformed, $R_3(r) = r$ and letting $R_2(r_h) = r_h$ gives the relation between $r_h$ and $r_c$. If there is external radial tension $T_{ext}$, the boundary condition $T_r = T_{ext}$ at $r = r_m$ gives the relation between $r_h$ and $r_c$. For our examples below, we have chosen $r_m = 1.2$ somewhat arbitrarily. In region 3, the radial tensions at two different points are related by

$$\frac{1}{T_r(r_1)} - \frac{2r_h^2}{r_1^2 - r_h^2} = \frac{1}{T_r(r_2)} - \frac{2r_h^2}{r_2^2 - r_h^2} \tag{13}$$

Fig. 6 shows the relation of the hole radius $R_h = R_2(r_h) = R_3(r_h)$ as a function of $r_c$ for various external tension $T_{ext}$. As $r_c$ increases, $R_h$ increases, *i.e.*, as the top rubber is less stretched, the radius of the hole becomes larger. In this graph, the curves end when the hole radius $R_h$ becomes equal to the coin radius, i.e., when $R_h = 1$. In practice, the invagination becomes unstable and the coin is released from the rubber before $R_h$ reaches $R_c$; the stability of the configuration is discussed in Sect. 2.1.2.

In the example shown in Fig 3, $r_c$ is about 0.5, and $R_h$ is about 0.77 according to the curve of $T_{ext} = 0$ in Fig. 6, in reasonable agreement with the hole size shown in Fig. 3. We note that if $r_c = 0$, then $R_h = r_h = 0$, and if $r_c = 1$, then $R_h = r_h = 1$.

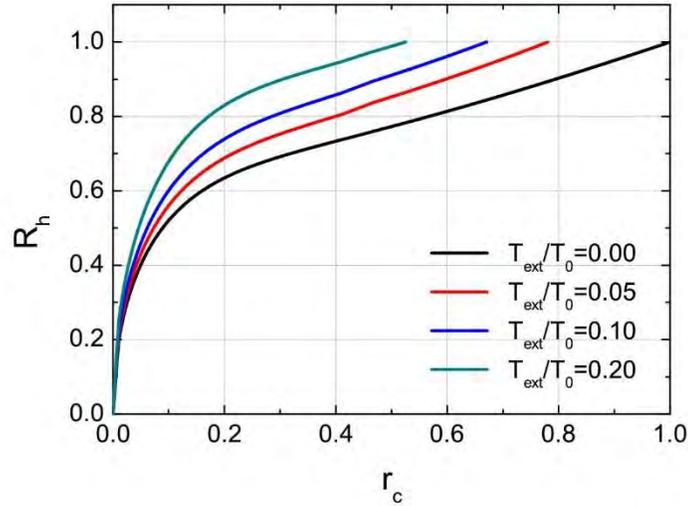

Fig. 6. Plot of hole radius $R_h$ vs. $r_c$ for different values of external tension. The lengths $R$ and $r$ are measured in units of the coin radius $R_c$.

Fig. 7 shows the solution $R(r)$ in all three regions for different applied tension at the edge. It is interesting to note that curves in the figure appear to be straight line segments, although the deformations in regions 2 and 3 are nonlinear functions of $r$. In region 1, $R' = 1/r_c > 0$, the rubber on the top of the coin is stretched uniformly, the slope of the line in region 1 indicates the stretch ratio. In region 2, $R' < 0$, indicating that the rubber has folded back, forming the pocket which holds the coin. In region 3, $R' > 0$, the rubber is folded forward. If there is no external tension, $T_{ext} = 0$, $R(r) = r$ is a straight line, and the rubber is undeformed. If there is external tension, $R(r)$ is a nonlinear function which depends on $T_{ext}$.

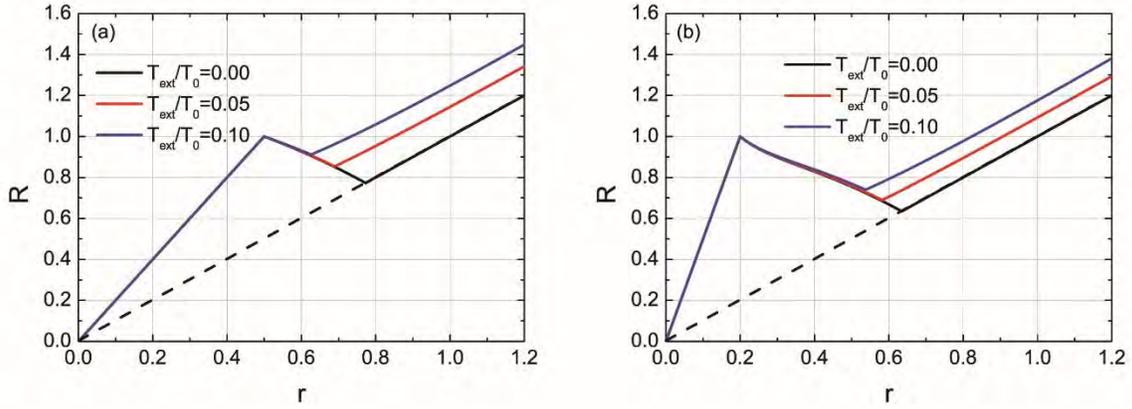

Fig. 7. Deformation $R(r)$ for stretch ratios (a) $r_c = 0.5$ and (b) $r_c = 0.2$ and various external tensions

The corresponding tensions per length, $T_r$ and $T_\phi$, are shown in Fig. 8. In region 1, both radial and tangential tensions are uniform and equal. In region 2, the radial tension $T_r$ decreases with $r$ and goes to zero at $r = r_h$, the boundary of region 2. In region 3, $T_r = T_\phi = 0$ if the external tension $T_{ext}$ is zero. If $T_{ext} > 0$, the radial tension $T_r$ increases to $T_{ext}$, while the tangential tension $T_\phi$ decreases with increasing $r$. In the limit that $r_m \to \infty$, $T_r = T_\phi$.

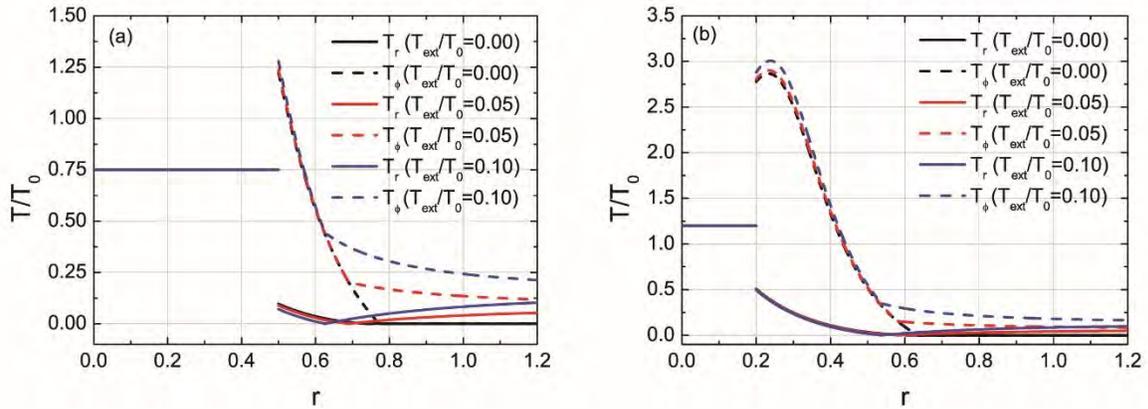

Fig. 8. Magnitudes of the radial and tangential tensions/length for stretch ratios (a) $r_c = 0.5$ and (b) $r_c = 0.2$ and various external tensions.

The thickness of the sheet, $h/h_0 = r/(sR'R)$, is shown in Fig. 9. In region 1, the thickness is uniform. If there is no external tension, then the thickness increases in region 2, and remains constant $h = h_0$ in region 3. When an external tension $T_{ext}$ is applied, the thickness increases with $T_{ext}$ in region 2, but decreases with $T_{ext}$ in region

3. The thickness appears to be discontinuous at the edge of the coin, but continuous at the hole. This follows from the continuity of $sR'$, which is continuous at the edge of the hole, but discontinuous at the edge of the coin.

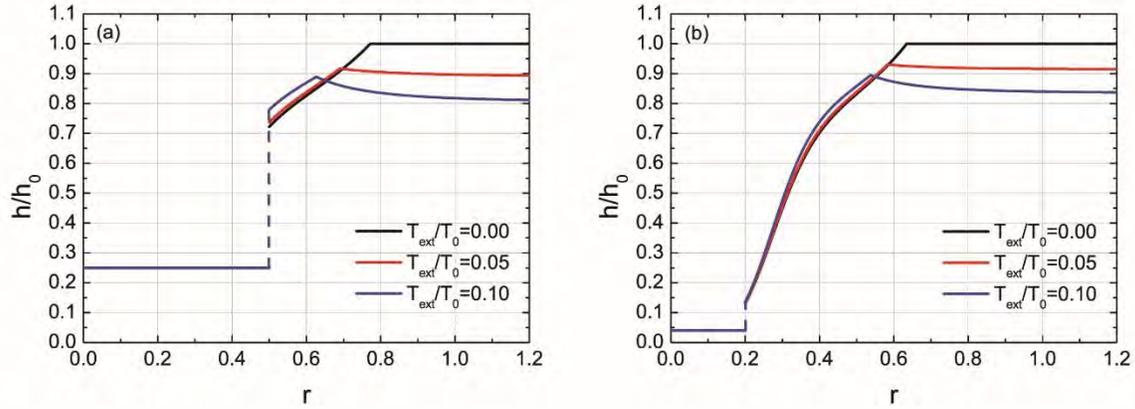

Fig. 9. Thickness $h/h_0$ of the rubber vs. $r$ for stretch ratio (a) $r_c = 0.5$ and (b) $r_c = 0.2$ and various external tension

### 2.1.2 Stability

As can be seen in Fig. 8, there are large differences in both radial and tangential tensions on the top, region 1, and bottom, region 2, of the coin at $r = r_c$. Whereas in planar regions of rubber, tangential hoop stresses compensate for changes in the radial stress and provide force balance, this is not the case at the edge of the coin, where tangential tension is balanced by force from the edge of the coin. The difference in the radial tension on top and bottom of the coin produces a net force, which must be balanced by friction of the rubber against the edge of the coin if the invagination is to be stable and the rubber is not to slip. A schematic is shown in Fig. 10.

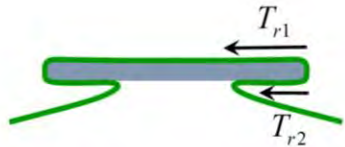

Fig. 10. Schematic of forces acting on the rubber at the edge of the coin

The radial tension per length, from Eq. (9), on top of the coin, in region 1 at $r = r_c$, is

$$T_{r1} = \frac{3}{2}(1 - r_c), \tag{14}$$

and below the coin, in region 2, at $r = r_c$, it is

$$T_{r2} = -\frac{3}{2} \frac{1 - \frac{r_h^2}{r_c^2} + \frac{r_h^2}{r_c^2} \ln \frac{r_h^2}{r_c^2}}{3 + \frac{r_h^2}{r_c^2}}. \tag{15}$$

To provide force balance, the friction force must equal the difference between the radial tensions on top and bottom at the edge of the coin. As the sheet passes around the edge of the coin from top to bottom, the normal force decreases, as in the case of a flexible rope passing around a stationary capstan shown in Fig. 11. Here the tension changes due to friction as the rope winds around the capstan, the change in tension is just $dT = \mu T d\theta$ where $\mu$ is the coefficient of static friction. The contribution of friction to radial tension can be estimated from Euler's equation, which gives the ratio of tensions in a rope or sheet winding through an angle $\theta$ around a capstan,

$$T_{r1} = T_{r2} e^{\theta \mu}. \tag{16}$$

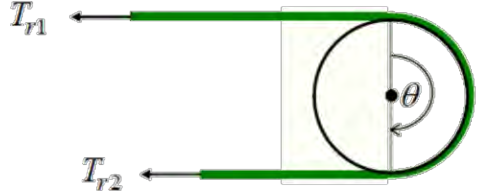

Fig. 11. Schematic of flexible rope passing around a stationary capstan.

(We have neglected here the contribution of the tangential tension along the edge of the coin to the normal force, since the coin is thin; curvature of the edge associated with thickness is much greater than curvature associated with the coin radius). In our case, $\theta = \pi$, and we must have, for stability, that

$$e^{\pi \mu} \geq \frac{T_{r1}}{T_{r2}}. \tag{17}$$

On substituting the expressions for the tensions from Eqs. (14) and (15), the critical friction coefficient $\mu_c$ is given by

$$\mu_c = \frac{1}{\pi} \ln\left(\frac{(1-r_c)(3r_c^2 + r_h^2)}{r_h^2 - r_c^2 - r_c r_h^2 \ln \frac{r_h^2}{r_c^2}}\right). \tag{18}$$

This is our general stability criterion. Fig. 12 shows the function $\mu_c(r_h, r_c)$ given by Eq. (17). Note that $r_h$ is a measure of the strength of the tension $T_{ext}$ applied at the edge of the rubber sheet. When $T_{ext}$ is increased, $r_h$ decreases and $\mu_c$ increases. The coin is released from the rubber when $\mu_c$ becomes equal to the friction coefficient of the rubber. If $r_c = 0.5$, and no external forces are applied at the edges, we have $r_h = 0.7727$, and $\mu_c = 0.6508$, i.e., the rubber stretched by the factor $1/r_c = 2$ can hold the coin if its friction coefficient is greater than 0.6508. As external tension is applied at the edges, $T_{r2}$ decreases, as can be seen in Fig. 8, $r_h$ increases, and the critical friction coefficient $\mu_c$ increases. For example, if $r_c = 0.5$, and $r_h = 0.6$, then $\mu_c = 0.804$. Parts of the curves shown in Fig. 12 are drawn dashed, indicating that the curves in this region are not physically realizable. In this parameter range, the E-L equation (7) does not have a solution which satisfies the condition $R' < 0$ in region 2. The region where the Hookean representation breaks down is discussed further below.

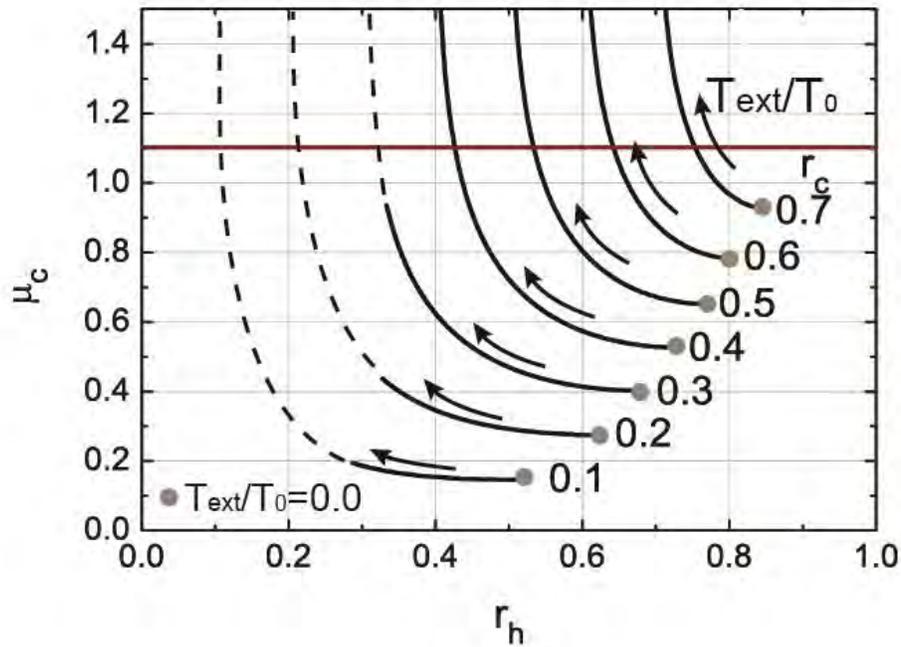

Fig. 12. The minimum friction coefficient $\mu_c$ which gives the configuration shown in Fig. 5 is plotted as a function of $r_h$ and $r_c$ from Eq. (17).

In the CTRT, the range of stretches is $0.2 < r_c < 0.5$. The coefficient of friction for latex on aluminum has been reported [16] to be as high as $\mu_c = 1.1$. Pushing down on the coin, as shown in Fig. 2 b., is equivalent to applying and increasing external tension at the edges of the rubber sheet. When the resulting critical friction coefficient exceeds the actual one, the invagination loses stability. Here is where the magic occurs: the sheet moves around the edge of the coin, region 2 disappears, and the coin is ejected from the invagination towards the bottom of the container.

As the critical tension is reached, the rubber starts to slip around the edge of the coin. As the rubber slips, $r_c$ increases. As shown in Fig. 17, as rc increases, even greater friction coefficient is required to stabilize the sheet than previously. Furthermore, as the rubber starts to slip, static friction between the rubber and the coin is replaced by (smaller) kinetic friction, which further contributes to the loss of stability. The loss of stability is thus subcritical, once the rubber starts to slip, it becomes increasingly unstable; the coin is ejected, and the rubber relaxes to the unfolded configuration.

### 2.1.3 Failure of the Hookean Description

Although the Hookean model generally describes the deformations of the rubber sheet surprisingly well, it does break down in certain regions of parameter space. Consider the radial tension in region 2 in the vicinity of $r \simeq r_c$ in the case of large external tension, when $r_h \simeq r_c$. The radial tension, from Eq. (9), is

$$T_r = T_o \frac{1}{2}\frac{r}{R}(2sR' + \frac{R}{r} - 3). \qquad (19)$$

Near $r \simeq r_c$, $R/r \simeq 1/r_c$. Since the radial tension has to vanish at $r = r_h$, we must have $2sR' \simeq 3 - \frac{1}{r_c}$. Clearly, this is not possible if $r_c < 1/3$. We see therefore that the Hookean description breaks down in the case of large strain ($r_c < 1/3$) and large external tension, where $r_h \simeq r_c$ which is outside of the region of validity of the linear approximation of the nonlinear energy density.

### 2.1.4 Discussion

The simple Hookean model above describes well the elastically stabilized invagination forming the 'pocket' which holds the coin and makes possible the magic. A key aspect of the configuration is the folding of the rubber sheet; it is interesting to consider how it comes about. First, it is not necessary for the sheet to fold at the edge of the coin at $r = r_c$; in the region $r_c < r < r_m$, the sheet could remain unfolded with $R' \geq 0$, satisfying the E-L equation with the deformation

$$R = (\frac{r_m^2 + r_c}{r_m^2 + r_c^2})r + (\frac{r_m^2 r_c(1-r_c)}{r_m^2 + r_c^2})\frac{1}{r}, \qquad (20)$$

which satisfies the boundary conditions $R(r_c) = 1$ and $T_r(r_m) = 0$. For this solution, $R' \leq 1$ for $r > r_c$, indicating compression rather than tension, and the free energy is greater than that of the solution corresponding to the folded invagination. The situation is similar under the coin. If the sheet folds at $r = r_c$, it is not necessary to fold again at the hole $r = r_h$. It could remain unfolded, but again the deformation for $r > r_h$ would be a compression with higher energy than the solution with the fold at $r = r_h$. The sheet thus folds where necessary to avoid compression yet satisfy the necessary boundary conditions.

## 2.2 The Neo-Hookean Model

A more accurate description of what happens is given by the neo-Hookean model of elasticity, which is better suited to large deformations than the Hookean model predicated on small strains. It is a special case of Rivlin's general model in Eq. (1) where $C_{10} = \frac{1}{2}G$ and all the other coefficients are zero. Assuming again the deformation to be of the form $\mathbf{R} = (R_r(r), 0, R_z(r,z))$ and that $\partial R_z / \partial r$ is small, the energy is given by

$$F = \frac{1}{2}G 2\pi h_0 \int_0^{r_m} [(R')^2 + (\frac{R}{r})^2 + (\frac{r}{RR'})^2 - 3] r \, dr. \tag{21}$$

The E-L equation

$$rR'' = \frac{1 + \frac{3r^3}{R^3 R'^3}}{1 + \frac{3r^2}{R^2 R'^4}} (\frac{R}{r} - R') \tag{22}$$

is now nonlinear. The radial and tangential tensions per unit length are

$$T_r = T_0 \frac{1}{4} \frac{r}{RR'} (R'^2 - \frac{r^2}{R^2 R'^2}), \tag{23}$$

and

$$T_\phi = T_0 \frac{1}{4} \frac{r}{RR'} (\frac{R^2}{r^2} - \frac{r^2}{R^2 R'^2}). \tag{24}$$

Unlike in the linear Hookean model, here in regions 2 and 3, the E-L equation cannot be solved analytically; the solution must be obtained numerically. The boundary conditions are the same as in the Hookean case. The numerical method we use here for the solution is the shooting method. Results of the numerical calculations are shown below.

Figs 13 and 14 allow comparison of the deformation $R(r)$ from the Hookean and the neo-Hookean models for different values of $r_c$. The two models gives qualitatively similar results: in Fig. 13 (a), they are remarkably close to each other. Fig 15 allows comparison of the predicted hole radius $R_h$ as function of $r_c$. Again, the two models give similar results; the difference is less than 20%.

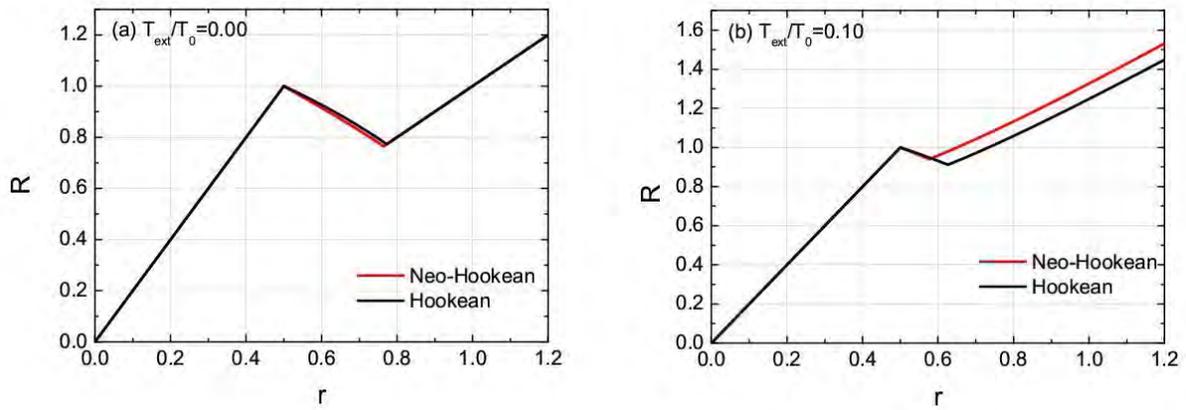

Fig. 13. Predictions of Hookean and Neo-Hookean models of the deformation $R(r)$ with $r_c = 0.5$ (a) $T_{ext} = 0$, (b) $T_{ext} = 0.10 T_0$.

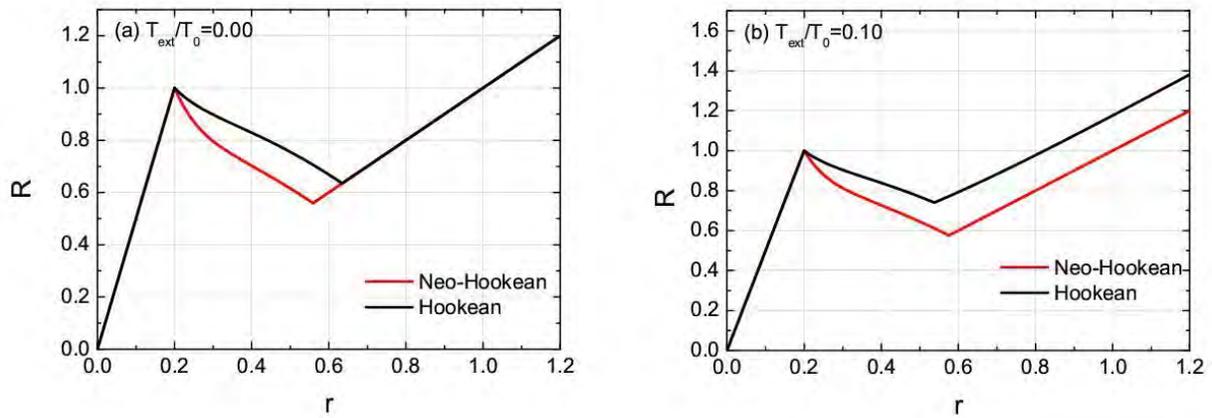

Fig. 14. Predictions of Hookean and Neo-Hookean models of the deformation $R(r)$ with $r_c = 0.2$ (a) $T_{ext} = 0$, (b) $T_{ext} = 0.10 T_0$.

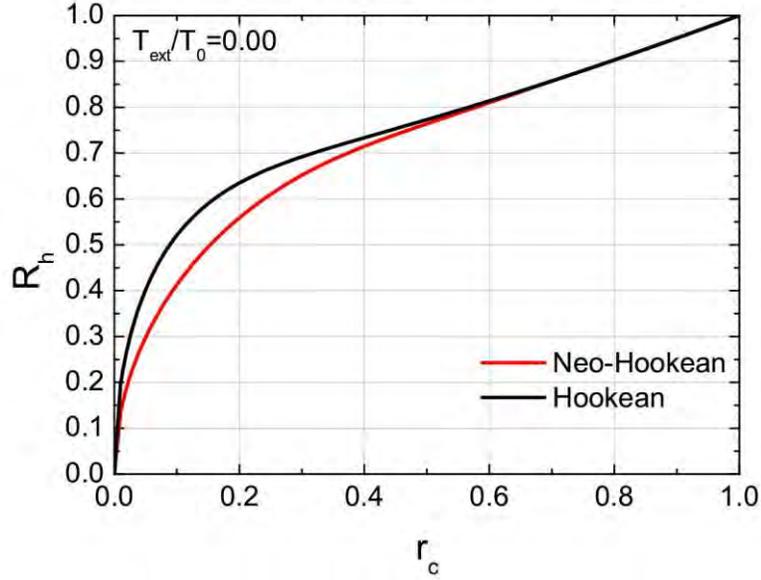

Fig. 15 Predictions of Hookean and Neo-Hookean models for the hole radius $R_h$ as function of $r_c$.

As in the case of the Hookean model, the invagination remains stable so long as

$$\mu \geq \frac{1}{\pi} \ln\left(\frac{T(r_{c-})}{T(r_{c+})}\right). \tag{25}$$

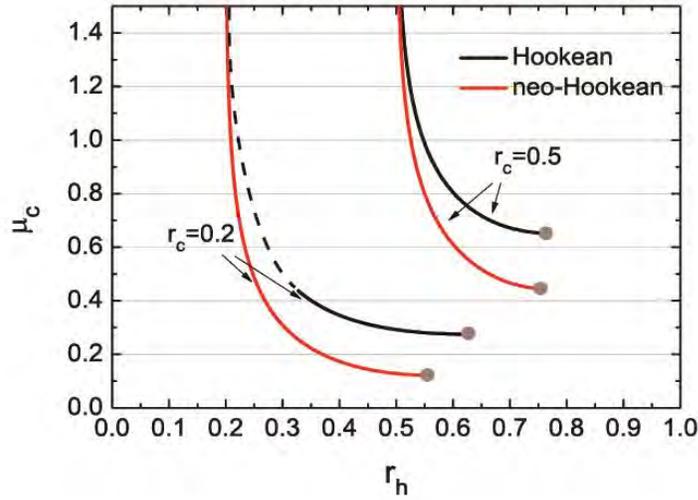

Fig. 16. Predictions of Hookean and neo-Hookean models for the critical friction coefficient as function of $r_h$.

Fig. 16. shows the critical friction coefficient $\mu_c$ as function of $r_h$ for both Hookean and neo-Hookean descriptions; it may be compared with Fig. 12. We note that the neo-Hookean model can predict the critical friction coefficient $\mu_c$ for arbitrary parameters $r_c$ and $r_h$ without difficulty, unlike the Hookean model.

Fig 17 shows the hole radius $R_h$ as function of $r_c$ for different values of external tension $T_{ext}$. For given value of $r_c$, the hole radius $R_h$ increases with $T_{ext}$, but cannot increase beyond the coin radius $R_c$. In practice, the maximum tension is determined by friction at the edge of the coin.

The stability limit determined by Eq. (25) is indicated by the dashed curves in Fig. 17. For given values of the friction coefficient $\mu$ and given external tension, the coin is released from the rubber when the solid line and dashed curve intersect.

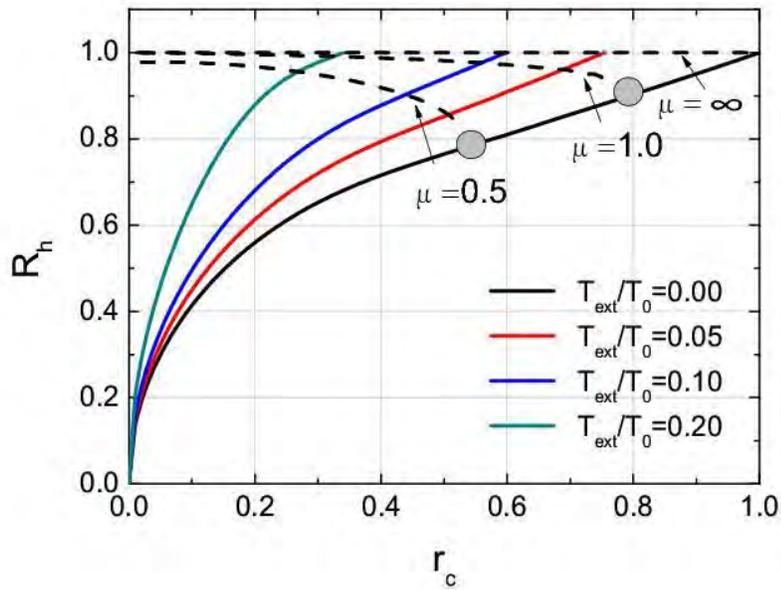

Fig. 17 The hole radius $R_h$ as function of $r_c$ for different values of external tension $T_{ext}$ in the neo-Hookean model. The solid curves indicate $R_h$ vs $r_c$; the dashed curves indicate the stability limit.

## Summary

In this paper we presented a simple analysis of the elastic deformation which underlies the CTRT where a coin, apparently on top of a rubber sheet covering a cup, appears to go through the rubber and fall into the cup. The coin, in fact always under the rubber sheet, is held in an invagination which is stabilized by elasticity and by friction; pushing down sufficiently hard on the coin (or equivalently, applying sufficient tension at the edges of the rubber) causes the invagination to become unstable, the coin is released, and falls to the bottom of the cup.

A simple volume conserving Hookean model for the cylindrically symmetric system provides two distinct linear Euler-Lagrange equations for different portions of the rubber which may be solved analytically. The solution indicates an invagination which holds the coin; stability of the invagination is provided by elasticity and friction between the rubber and the edge of the coin; the domain of stability can be determined via Euler's capstan equation.

To test the validity of the Hookean model, we have carried out a neo-Hookean analysis. This provides a non-linear Euler-Lagrange equation, which needs to be solved numerically. Predictions of the Hookean and neo-Hookean models were found to be in good agreement in a large region of parameter space; the failure and region of stability of the Hookean model were identified. Although we are unaware of similar invaginations, stabilized by elasticity and friction, occurring in nature, we suspect that they may exist in biology.

# Acknowledgements

We acknowledge useful discussions with David Andelman, John Ball and Epifanio Virga. M. D. acknowledges the financial support of the Chinese Central Government in the program of "Thousand talents" and the NSFC grant (No. 21434001). F. M. acknowledges the financial support of NSFC under grant numbers 31128004, 91027045, 10834014, 11175250, National Basic Research Program of China (973 program, No. 2013CB932800). P.P-M and X. Z. acknowledge support by the NSF under EFRI-1332271, IIP-1114332 and DMS-1212046

# Appendix: A Brief History of the CTRT.

The CTRT, or 'coin-through-the-rubber' trick was a favorite of Martin Gardner. One of us (P.P-M.) learned this trick from William G. Unruh, who learned it from Sir Roger Penrose, who learned it from Martin Gardner. Rudy Rucker mentions Gardner showing him the trick [17] and indicating that the trick was well known among close-up magicians. Rucker recorded Gardner's 'explanation' while performing the trick [18]:

*I want to move this dime through the membrane into the glass. How? I'll use the fourth dimension. If something travels through the fourth dimension it can go "around" a 3D obstacle.*

*Think of a square in Flatland. If I draw a cup with a lid in the plane, it's a single unbroken line. (Martin drew this shape on a piece of paper, like a U with a line across the top with a square beside it. And a square can't push something inside that line without crossing the edge. But if something rises out of the square's 2D space into the 3rd dimension it can move over the line and settle down inside it.*

*If we wall off a little volume of our space with a glass and with a rubber membrane, as I have done, we can't get inside that volume without breaking the glass or puncturing the membrane. In particular, I can't put this coin inside the glass. But if I can lift the coin up into the fourth dimension, I can move "over" the membrane and land inside the glass.*

*I'll let you help. It would be dangerous for you to push the coin through the fourth dimension with your bare finger. So I'll rest a quarter on top of the dime. And you push down on the quarter. And we'll focus on the fourth dimension while you do it. And the dime will slide through hyperspace and end up in the glass.*

*Ready? Push. (clatter of the dime in the glass). You see? The dime travelled through four dimensional space to get around the rubber membrane.*

The CTRT, also known as the 'dental dam trick' is credited [19] as a creation of Lubor Fiedler. Fiedler (1933-2014) was born in Brno, Czechoslovakia. He moved to Vienna, where was selling magic tricks to the public on the streets and at the entrances to supermarkets in order to make ends meet. The first description of the CTRT was published by him [20] in 1958. The trick was subsequently marketed by Gene Gordon without permission or credit as Dam Deception [21] in 1963. The dental dam, used in CTRT, was invented by Dr. Sandford Christie Barnum, in Monticello, New York in 1864 [22]. Rather than patenting it, Barnum presented it as a free gift to the dental profession. In Europe, the dental dam became known as the Kofferdam.

1. https://www.youtube.com/watch?v=64G2DjMwA-M
2. https://www.youtube.com/watch?v=3IB7wTKvGmw
3. Murat, D., Quinlan, A., Vali, H., Komeili, A.: Comprehensive genetic dissection of the magnetosome gene island reveals the step-wise assembly of a prokaryotic organelle. P Natl Acad Sci USA **107**(12), 5593-5598 (2010).
4. Itoh, T., Erdmann, K.S., Roux, A., Habermann, B., Werner, H., De Camilli, P.: Dynamin and the actin cytoskeleton cooperatively regulate plasma membrane invagination by BAR and F-BAR proteins. Dev Cell **9**(6), 791-804 (2005).
5. Sens, P., Turner, M.S.: Theoretical model for the formation of caveolae and similar membrane invaginations. Biophysical journal **86**(4), 2049-2057 (2004).
6. Ben-Dov, N., Korenstein, R.: Enhancement of cell membrane invaginations, vesiculation and uptake of macromolecules by protonation of the cell surface. PloS one **7**(4), e35204 (2012).
7. Wolff, J., Komura, S., Andelman, D.: Budding of domains in mixed bilayer membranes. Phys Rev E **91**(1) 012708-9 (2015).
8. Lipowsky, R.: Remodeling of membrane compartments: some consequences of membrane fluidity. Biol Chem **395**(3), 253-274 (2014).
9. McMahon, H.T., Boucrot, E.: Molecular mechanism and physiological functions of clathrin-mediated endocytosis. Nat Rev Mol Cell Bio **12**(8), 517-533 (2011).
10. Lewis, W.H.: Mechanics of invagination. The Anatomical record **97**(2), 139-156 (1947).
11. Rauzi, M., Hocevar Brezavscek, A., Ziherl, P., Leptin, M.: Physical models of mesoderm invagination in Drosophila embryo. Biophysical journal **105**(1), 3-10 (2013).
12. Ball, J.M.: Discontinuous Equilibrium Solutions and Cavitation in Non-Linear Elasticity. Philos T R Soc A **306**(1496), 557-611 (1982).

13. Bower, A.F.: Applied mechanics of solids. CRC Press, Boca Raton (2010)
14. Boyce, M.C., Arruda, E.M.: Constitutive models of rubber elasticity: A review. Rubber Chem Technol **73**(3), 504-523 (2000).
15. Audoly, B., Pomeau, Y.: Elasticity and geometry : from hair curls to the non-linear response of shells. Oxford University Press, Oxford ; New York (2010)
16. Hur, P., Motawar, B., Seo, N.J.: Hand breakaway strength model-Effects of glove use and handle shapes on a person's hand strength to hold onto handles to prevent fall from elevation. J Biomech **45**(6), 958-964 (2012).
17. http://martin-gardner.org/Testimonials.html
18. Rudy Rucker (priv. comm.)
19. http://geniimagazine.com/magicpedia/Lubor_Fiedler
20. Fiedler, L.: Eine fast unmögliche Münzenwanderung. Zauberkunst **4**(10), 129 (1958).
21. http://www.geniimagazine.com/magicpedia/Gene_Gordon
22. Koch, C.R.E., Thorpe, B.L.: History of dental surgery, vol. 2. The National Art Publishing Company, Chicago, Ill., (1909)